\documentclass{emulateapj}
\newcommand{\chandra}{\textit{Chandra}}
\newcommand{\cxo}{\textit{Chandra X-ray Observatory}}

\shorttitle{Most Luminous Supernova Ever}
\shortauthors{Smith et al.}
\begin{document}


\title{SN 2006gy: Discovery of the Most Luminous Supernova Ever
Recorded, Powered by the Death of an Extremely Massive Star like Eta
Carinae}

\author{Nathan Smith\altaffilmark{1}, Weidong Li\altaffilmark{1}, Ryan
J. Foley\altaffilmark{1}, J.\ Craig Wheeler\altaffilmark{2}, David
Pooley\altaffilmark{1,3}, Ryan Chornock\altaffilmark{1}, Alexei
V. Filippenko\altaffilmark{1}, Jeffrey M.\ Silverman\altaffilmark{1},
Robert Quimby\altaffilmark{2}, Joshua S. Bloom\altaffilmark{1}, and
Charles Hansen\altaffilmark{1} }

\altaffiltext{1}{Department of Astronomy, University of California, 
Berkeley, CA 94720-3411.}

\altaffiltext{2}{Department of Astronomy, University of Texas, 1
University Station C1400, Austin, TX 78712.}

\altaffiltext{3}{Chandra Fellow.}

\begin{abstract}

We report the discovery and early observations of the peculiar Type
IIn supernova (SN) 2006gy in NGC~1260, revealing that it reached a
peak visual magnitude of about $-$22, making it the most luminous supernova
ever recorded.  It had a very slow rise to maximum that took about
70~d and stayed brighter than $-$21 mag for about 100~d.  It is not
yet clear what powers the enormous luminosity and the total radiated
energy of $\sim 10^{51}$ erg, but we argue that any known mechanism
--- thermal emission, circumstellar interaction, or $^{56}$Ni decay
--- requires a very massive progenitor star.  The circumstellar
interaction hypothesis would require truly exceptional conditions
around the star, which, in the decades before its death, must have
experienced a luminous blue variable (LBV) eruption like the 19th
century eruption of $\eta$~Carinae.  However, this scenario fails to
explain the soft and unabsorbed X-ray emission detected by the {\it
Chandra X-ray Observatory}, which constrains the progenitor's
mass-loss rate to be three orders of magnitude too small.
Alternatively, radioactive decay of $^{56}$Ni may be a less
objectionable hypothesis, but it would imply a large Ni mass of
$\sim$22 M$_{\odot}$, requiring that SN~2006gy was a pair-instability
supernova where the star's core was obliterated (rather than forming
a neutron star or a black hole).  While this is still
uncertain, SN~2006gy is the first supernova for which we have good
reason to suspect a pair-instability explosion.  Independent of the
energy budget, a narrow H$\alpha$ emission line from unshocked
circumstellar gas also suggests a very massive progenitor star. The
shell has a mass of several M$_{\odot}$ of hydrogen and expansion
speeds of 130--260 km s$^{-1}$, ruling out progenitor stars with
initial masses below 40~M$_{\odot}$.  Based on a number of lines of
evidence, we eliminate the hypothesis that SN~2006gy was a ``Type IIa''
event --- that is, a white dwarf exploding inside a hydrogen envelope.
Instead, we propose that the progenitor may have been a very massive,
evolved object like $\eta$~Carinae that, contrary to expectations,
failed to completely shed its hydrogen envelope before it died.  Our
interpretation of SN~2006gy implies that some of the most massive
stars can explode prematurely during the LBV phase, preventing them
from ever becoming Wolf-Rayet stars.  SN~2006gy also suggests that
some of the most massive stars can create brilliant supernovae instead
of experiencing ignominious deaths through direct collapse to a black
hole. If such a fate is common among the most massive stars, then
observable supernovae from Population III stars in the early universe
will be more numerous than previously believed.

\end{abstract}


\section{INTRODUCTION}

Supernovae (SNe) resulting from the deaths of massive stars span a
wide range of peak absolute visual magnitude, typically between $-$15
and $-$20.5, and usually reaching their peak within about 20~d. They
also exhibit a range of spectral properties depending on the extent to
which products of nuclear burning are exposed at their surface, as
well as the expansion speed and the amount of circumstellar material.
Their diversity depends on the star's initial mass and rate of mass
loss during its lifetime.  Current expectations are that stars born
with initial masses above $\sim$40~M$_{\odot}$, which never become red
supergiants (RSGs; Humphreys \& Davidson 1979), will shed their
hydrogen envelopes to expose their He core before they die (e.g.,
Abbott \& Conti 1987).  As Wolf-Rayet (WR) stars, they are then
expected to explode, producing Type Ib/c SNe (see Filippenko 1997).
Based on observations of SN~2006gy that we discuss here, we speculate
that this scenario does not always apply.

One way to prevent a star from reaching the WR phase before explosion
would be if the star's mass-loss rate is insufficient to shed the
hydrogen envelope before the end of core He burning.  This is thought
to be the case for massive stars in the early universe, because their
much lower (or zero) metallicity should make their line-driven stellar
winds very inefficient (Baraffe et al.\ 2001; Kudritzki 2002; Heger et
al.\ 2003).  Depending on the mass at the time of death, very massive
stars in this predicament might suffer a pair-production instability
explosion (Barkat et al.\ 1967; Fraley 1968; Bond, Arnett, \& Carr
1984; Heger \& Woosley 2002), where the star's core is obliterated
instead of collapsing to a black hole.

However, there are reasons to suspect that the mass-loss properties of
stars in the local universe may not be so different from these early
stars.  Namely, recent studies of line-driven winds from O-type stars
and WR stars have shown that their winds are highly clumped, requiring
that their mass-loss rates through line-driven winds on the main
sequence could be an order of magnitude lower than previously believed
(Fullerton et al.\ 2006; Bouret et al.\ 2005).  In that case, for
stars with initial masses above $\sim$40~M$_{\odot}$ that never become
RSGs, the burden of mass loss falls to the post-main-sequence luminous
blue variable (LBV) phase, when very massive stars suffer multiple
giant eruptions that shed several M$_{\odot}$ in just a few years
(Smith \& Owocki 2006). If these LBV eruptions are not sufficient to
remove the star's entire outer hydrogen envelope fast enough, as may
be the case for the most massive stars above 100~M$_{\odot}$, then the
star would seem to explode early as an LBV producing a Type IIn event.
Interestingly, Gal-Yam et al.\ (2007) find that the rate of Type IIn
events is in broad agreement with the hypothesis that they are the
explosions of extreme LBVs.  The fact that giant LBV eruptions are
continuum driven may hint that low-metallicity stars may be capable of
shedding mass after all (Smith \& Owocki 2006), which would affect the
range of initial masses that are subject to the pair instability in
Population III stars.  Because stars that begin their lives above
100~M$_{\odot}$ are so few in number, their end fates are poorly
constrained by observations (see Gal-Yam et al.\ 2007 for relevant
discussion), and are still an open question.  For these reasons, any
potential detection of a pair-instability supernova in the modern
universe would be of great interest to stellar astrophysics.  Here we
explore this notion, along with others, as a possible explanation for
the bizarre properties of SN~2006gy.

SN~2006gy in the peculiar S0/Sa galaxy NGC~1260 was discovered and
confirmed by the Texas Supernova Search (TSS; Quimby 2006a) with the
ROTSE~IIIb telescope (Akerlof et al.\ 2003) at McDonald Observatory in
unfiltered images (Quimby 2006b) taken on 2006 Sep. 18.3 (UT dates are
used throughout this paper). It was initially classified (Harutyunyan
et al. 2006) as a SN~II (actually SN~IIn, based on the written
description), but Prieto et al. (2006) nearly simultaneously suggested
that the object was instead a bright active galactic nucleus
(AGN). However, in the subsequent month, our group continued to follow
SN~2006gy, and with additional astrometric, photometric, and
spectroscopic data we announced that it did indeed appear to be a SN
after all, and not an AGN (Foley et al.\ 2006).  In this paper we
present additional data and analysis of SN~2006gy, leading us to
propose that it marked the death of a very massive star with much of
its hydrogen envelope still intact, while surrounded by a massive
circumstellar nebula.  In many respects, the type of progenitor we
infer for SN~2006gy resembles the LBV star $\eta$~Carinae in our own
Galaxy, as discussed below.

\section{OBSERVATIONS}

\subsection{Imaging and Photometry}

Figure 1 shows a laser guide star (LGS) adaptive optics (AO)
near-infrared image of SN~2006gy and the nucleus of its host galaxy
NGC 1260, revealing a clear offset of the SN from the galaxy
center. Images at three wavebands ($J$, $H$, and $K_s$) were obtained
on 2006 Nov. 4 using the AO system in LGS mode (Lloyd et al.\ 2000;
Max et al.\ 1997) on the Shane 3-m telescope at Lick Observatory. The
total integration time in each band was 480~s, accumulated over 8
exposures. The native scale of the $256 \times 256$ pixel Rockwell
PICNIC array is 0$\farcs$076 pixel$^{-1}$ (Perrin 2007).  Mosaiced
images have a scale of 0$\farcs$04 pixel$^{-1}$. The SN itself was
bright enough to use as a ``tip-tilt'' star for the LGS system. The
effective resolution (full width at half-maximum intensity; FWHM) is
0$\farcs$2 in the $H$ band. The measured offset of the SN from the
centroid of the galactic nucleus is 0$\farcs$941 west, 0$\farcs$363
north, with a 1$\sigma$ uncertainty of 0$\farcs$01 in each direction;
this confirms and improves the earlier offset measurement (Foley et
al.\ 2006) of 0$\farcs$880 west, 0$\farcs$140 north, $\pm$0$\farcs$08.
SN~2006gy is therefore located about 350 pc from the galaxy's center
(at its assumed distance of $\sim$73 Mpc), confirming that it is not
an AGN.\footnote{Ironically, NGC~1260 may contain a faint AGN after
all, although SN~2006gy is a real SN explosion.  Later in this paper
we also present an X-ray image of SN~2006gy which shows two sources,
one being the SN and the other the nucleus of NGC~1260.}

Figure 2 shows the $R$-band light curve of SN~2006gy obtained by our
group using the Katzman Automatic Imaging Telescope (KAIT; Filippenko
2003) at Lick Observatory, compared to a sample of several other
representative SN light curves.  The unfiltered KAIT images for SN
2006gy were used to derive an $R$-band light curve.  As demonstrated
by Riess et al.\ (1999) and Li et al.\ (2003), the best match to
broad-band filters for the KAIT unfiltered data is the $R$ band. Each
image is aligned to a deep pre-SN image, and the contamination of the
host-galaxy emission is carefully removed.  The net flux for the SN is
then compared to 19 bright stars using calibrations from the USNO~B1
catalog.  We list the KAIT apparent $R$ magnitudes of SN~2006gy in
Table 1. To put the flux of SN~2006gy on an absolute magnitude scale,
we adopt a distance to the host galaxy NGC 1260 of 73.1 Mpc, using
$H_0 = 72$ km s$^{-1}$ Mpc$^{-1}$ and using a recession velocity for
the central cluster galaxy of 5361 km~s$^{-1}$. We also assume a
Galactic reddening of $A_R = 0.43$ mag (Schlegel et al.\ 1998) and a
host-galaxy reddening of $A_R = 1.25 \pm 0.25$ mag (see \S 2.2 and
Fig.\ 3). In Figure 2 we plot days since explosion instead of days
since discovery.  Our first measurement with KAIT was a non-detection
made on 2006 Aug. 26, which was 23~d before the discovery of SN 2006gy.
Judging from the slowly rising curve, we estimate that the explosion
date was roughly 6~d before the KAIT non-detection.

\begin{deluxetable}{lcc}
\tabletypesize{\scriptsize}
\tighten
\tablewidth{2.2in}
\tablecaption{KAIT Photometry of SN~2006gy}
\tablehead{
  \colhead{MJD} &\colhead{$m_R$} &\colhead{Err.}
}
\startdata
3973.96 &(18.62) &0.03 \\
3982.00 &16.22 &0.03 \\
3987.98 &15.72 &0.03 \\
3995.04 &15.12 &0.03 \\
4003.03 &14.72 &0.03 \\
4007.95 &14.62 &0.03 \\
4014.97 &14.42 &0.03 \\
4020.99 &14.32 &0.03 \\
4026.92 &14.27 &0.03 \\
4033.92 &14.22 &0.03 \\
4038.85 &14.22 &0.03 \\
4047.81 &14.28 &0.03 \\
4049.92 &14.28 &0.03 \\
4055.87 &14.38 &0.03 \\
4061.88 &14.49 &0.03 \\
4068.89 &14.60 &0.03 \\
4076.83 &14.90 &0.03 \\
4087.75 &15.15 &0.03 \\
4089.77 &15.24 &0.03 \\
4092.75 &15.26 &0.03 \\
4094.76 &15.46 &0.03 \\
4098.76 &15.45 &0.03 \\
4102.74 &15.54 &0.03 \\
4106.71 &15.71 &0.03 \\
4121.71 &15.97 &0.03 \\
4125.72 &16.03 &0.03 \\
4130.60 &16.26 &0.03 \\
4133.64 &16.29 &0.03 \\
4134.63 &16.24 &0.03 \\
4135.61 &16.38 &0.05 \\
4137.69 &16.35 &0.04 \\
4150.62 &16.58 &0.03 \\
4162.65 &16.68 &0.05 \\
4166.63 &16.76 &0.05 \\
4168.64 &16.71 &0.05 \\
4170.63 &16.70 &0.05 \\
4171.63 &16.72 &0.05 \\
4173.63 &16.76 &0.06 \\
4174.64 &16.75 &0.07 \\
4175.64 &16.59 &0.05 \\
4177.64 &16.79 &0.05 \\
4178.64 &16.77 &0.05 \\
4181.64 &16.71 &0.05 \\
4183.64 &16.74 &0.05 \\
4184.64 &16.74 &0.08 \\
\enddata
\end{deluxetable}

\begin{figure}
\epsscale{0.93}
\plotone{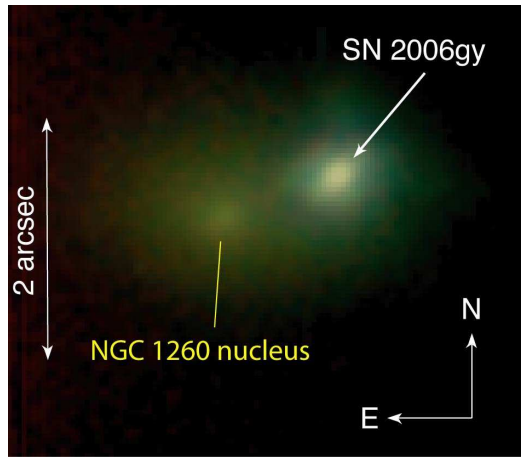}
\caption{Laser guide star adaptive optics image of SN~2006gy and the
nucleus of NGC 1260, showing a clear offset of the SN from the galaxy
center.  Blue is $J$ band (1.25 $\mu$m), green is $H$ band (1.65
$\mu$m), and red is $K_s$ band (2.2 $\mu$m).}
\end{figure}

\begin{figure}
\epsscale{1.15}
\plotone{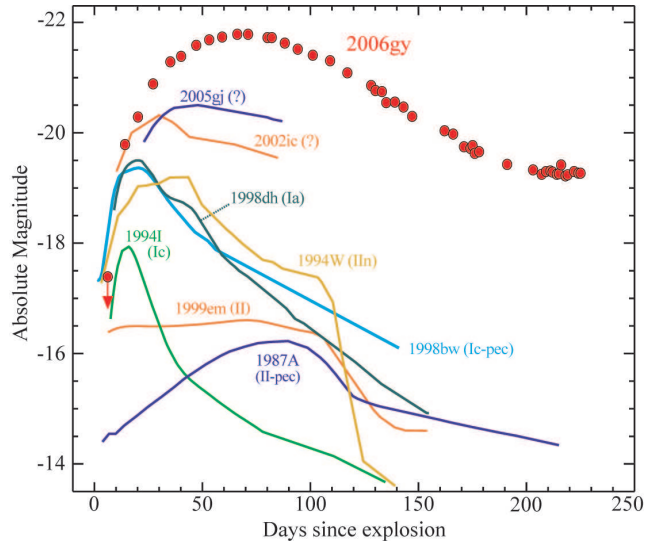}
\caption{Comparison of the absolute $R$-band light curve of SN~2006gy
with those of other SNe. We plot days since explosion, which we judge to
be $\sim$29~d prior to the discovery of SN~2006gy.  SN 1998dh is a
typical SN~Ia, and the data are from our unpublished photometric
database, with a typical absolute magnitude of $M_R = -19.5$ mag
assumed. SN~1999em is a typical Type~II (Leonard et al.\ 2002),
SN~1994I is a well-observed SN~Ic (Richmond et al.\ 1996), and
SN~1998bw is a peculiar SN~Ic (Galama et al.\ 1998).  SN~1987A is a
peculiar SN~II, with a broad light curve but a low luminosity (from
Hamuy et al.\ 1990). SN~1994W is a SN~IIn that is powered by strong
interaction with its circumstellar material (Sollerman et al.\ 1998).  
We also plot two unusual SNe that are relevant to the discussion of 
SN~2006gy: SN 2002ic (Hamuy et al.\ 2003) and SN 2005gj (Aldering 
et al.\ 2006).}
\end{figure}

\subsection{Lick and Keck Spectroscopy}

\begin{figure*}
\epsscale{0.6}
\plotone{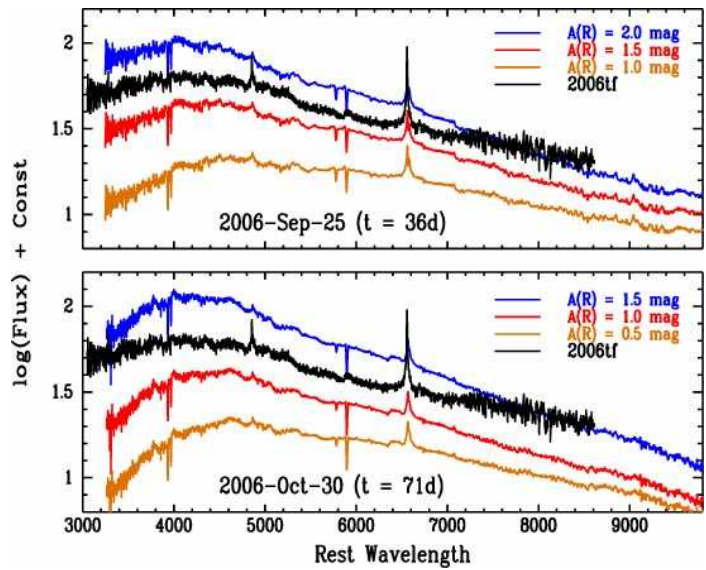}
\caption{Lick Observatory spectra of SN~2006gy at two different
  epochs, corrected for a range of assumed host-galaxy reddening
  corresponding to the values of $A_R$ listed at right (Cardelli et
  al.\ 1989).  This extinction is in addition to Galactic extinction
  of $A_R = 0.43$ mag.  These are compared to the day 32 spectrum of the
  Type IIn SN~2006tf (black) from our database, which is a SN with a
  spectrum similar to that of SN~2006gy, but seems to show little reddening.
  We adopt $A_R = 1.25 \pm 0.25$ mag for SN~2006gy; see text.}
\label{fig:red}
\end{figure*}

Figure~\ref{fig:red} shows two visual-wavelength spectra of SN~2006gy
obtained on 2006 Sep.\ 25.5 and 2006 Oct.\ 30.4 using the Kast double
spectrograph (Miller \& Stone 1993) mounted on the Lick Observatory
3-m Shane telescope. The long slit of width 2$\arcsec$ was aligned
along the parallactic angle (Filippenko 1982). The data were reduced
using standard techniques as described by Foley et al.\ (2003) and
references therein.  The spectra were corrected for atmospheric
extinction (Bessell 1999; Matheson et al.\ 2000) and then flux
calibrated using standard stars observed at an airmass similar to
that of the SN.

The closest match to SN~2006gy in our spectral database is SN~2006tf,
taken on 2007 Jan.\ 13, as shown in Figure~\ref{fig:red}.\footnote{SN
2006tf was discovered in the course of the Texas Supernova Search on
2006 Dec. 12 UT (Quimby et al.\ 2007). With a discovery magnitude of
16.7 and a redshift $z = 0.074$, the SN has an absolute magnitude of
$-$20.7, which is extremely luminous but still less so than SN
2006gy. Our follow-up photometry also suggests that SN~2006tf exhibits
a light-curve shape similar to that of SN~2006gy.}  The red continuum
shape of SN~2006gy is unusual for SNe~IIn, which are typically much
bluer (Schlegel 1990), so we have plotted the SN~2006gy spectrum after
removal of various amounts of reddening for comparison. Although a
direct comparison to SN~2006tf is complicated by the temporal evolution,
the early (day 36) spectrum of SN~2006gy seems most consistent with
$A_R = 1.5$ mag, while the later (day 71) spectrum is more consistent
with $A_R = 1.0$ mag (the spectra of SN~2006gy were already corrected
for Galactic extinction of $A_R = 0.43$ mag, as noted earlier).
Comparison with other SNe~IIn at similar phases (not shown) also
suggests values of the host-galaxy value of $A_R = 1.0$ to 1.5 mag.  
We therefore adopt $A_R = 1.25 \pm 0.25$ mag for SN~2006gy.
The extinction could be higher if SN~2006tf has its own significant
reddening, although it appears to have very weak Na~{\sc i}~D
absorption.  The strong Na~{\sc i}~D absorption in the spectrum of
SN~2006gy may suggest higher reddening than we have assumed here, so
our estimates of luminosity for SN~2006gy are conservative.

\begin{figure*}
\epsscale{0.9}
\plotone{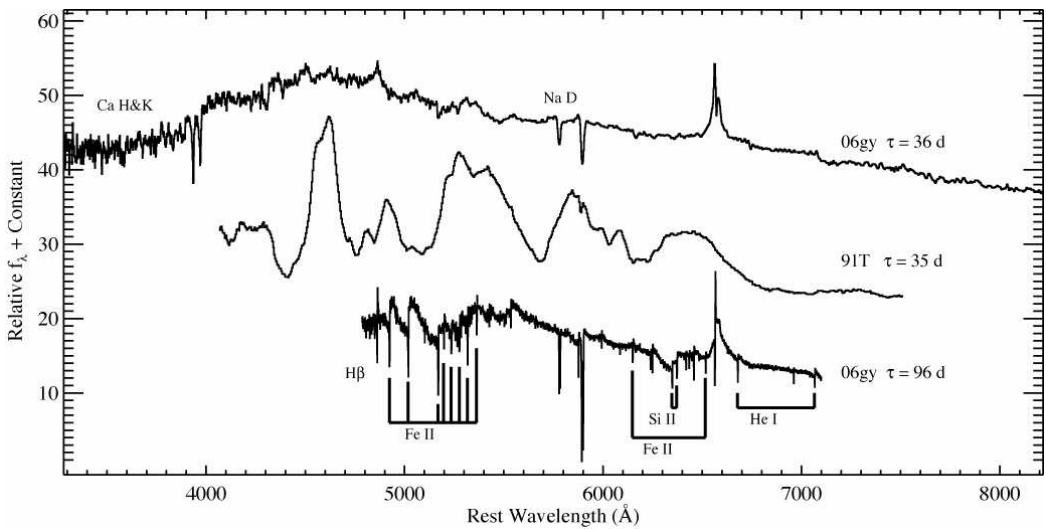}
\caption{Dereddened visual-wavelength spectra of SN~2006gy at $t=36$~d
and 96~d after explosion, obtained at Lick Observatory and with the
Keck~II telescope, respectively.  Several narrow absorption lines in
our high-resolution Keck spectrum have been marked, but there are some
remaining unidentified lines.  Also plotted is a spectrum of the Type
Ia SN~1991T at $t=35$~d (Filippenko et al.\ 1992) for comparison with
our day 36 spectrum of SN~2006gy; there is essentially no similarity
between the two spectra.}
\label{fig:lick}
\end{figure*}

Figure~\ref{fig:lick} shows the day 36 Lick spectrum from
Figure~\ref{fig:red}, and also a spectrum with a smaller wavelength
range and higher spectral resolution of $R \approx 4500$ taken near
maximum light on day 96.  The latter spectrum was obtained on 2006
Nov.\ 24.51 using the DEIMOS spectrograph (Faber et al.\ 2003) on the
Keck~II telescope. Using a customized version of the DEEP data
reduction pipeline, we obtained sky-subtracted, rectified
two-dimensional images, and wavelengths were calibrated with respect
to an internal calibration lamp (Foley et al.\ 2007).  We checked carefully 
to make sure that the sky-subtraction procedure did not artificially
introduce narrow absorption components; this is implausible based on
the final results anyway, since H and He~{\sc i} lines show similar
blueshifted absorption profiles.  We corrected for telluric absorption
(Matheson et al. 2000) by comparison with the standard star BD+284211.

Figure~\ref{fig:ha} shows the H$\alpha$ profile of SN~2006gy near
maximum light from a portion of the same Keck spectrum in Fig.\
\ref{fig:lick}, with the flux normalized to the underlying continuum
level, and the velocity scale chosen with the narrow H$\alpha$
emission feature at $v=0$ km s$^{-1}$.  The H$\alpha$ profile in
Figure~\ref{fig:ha} reveals several different characteristic
velocities relevant to interpretations of SN~2006gy.  First, the very
narrow emission component (FWHM $\approx$ 100 km s$^{-1}$) has an
associated P Cygni absorption feature that indicates outflow speeds of
130 km s$^{-1}$ (the trough) to 260 km s$^{-1}$ (the blue edge) in the
unshocked circumstellar gas.  In addition to H$\alpha$, several lines
identified in Figs~\ref{fig:lick} and \ref{fig:ha} also have narrow
absorption features.

A broad H$\alpha$ emission component has an apparent FWHM $\approx$
2400 km s$^{-1}$ that is similar to H$\beta$ at early times
(Harutyunyan et al.\ 2006).  The true unabsorbed FWHM of this broad
H$\alpha$ component is larger because of the broad blueshifted
absorption.  Extended faint wings out to $\pm$6,000 km s$^{-1}$ may be
caused either by electron scattering or by the fastest SN ejecta.

The blue edge of the broad, blueshifted H$\alpha$ absorption in
Figure~\ref{fig:ha} indicates an outflow speed of 4,000 km s$^{-1}$,
where the emission jumps back up just to the level that would be
expected for a symmetric profile.  This jump is readily apparent when
we take the redshifted side of the broad emission profile and reflect
it to the blue side, to simulate what a symmetric profile would look
like (Fig.\ \ref{fig:ha}).  Because this absorption traces the speed
of the dominant absorbing material along the line of sight at this
epoch, we take this speed of 4,000 km s$^{-1}$ to represent dense
material swept up by the SN blast wave in the circumstellar material
(CSM) interaction hypothesis, which should closely trace the speed of
the blast wave itself.

The broad-line profile differs from the smooth broad parts of
H$\alpha$ profiles normally seen in SNe~IIn (e.g., Chugai et al.\
2004).  The blueshifted absorption trough flattens out and does not
descend below the underlying continuum level.  This may hint that the
continuum luminosity and H$\alpha$ emission/absorption have different
origins, and provides important clues to the shell optical depth and
CSM density.  For example, the blueshifted absorption may arise in
shocked CSM gas, whereas the continuum luminosity may originate in the
SN ejecta.  Asymmetric geometry in the CSM obviously may be relevant.
These details have some bearing on the hypotheses for the power
sources discussed in \S 3.3 and \S 3.4.  In any case, this broad,
blueshifted H$\alpha$ absorption probably shares an origin with the
broad, blueshifted absorption features for other lines identified in
Figure~\ref{fig:lick}.  In light of possible geometric complexities,
we defer a detailed discussion of the line profiles to a later paper.

\subsection{X-ray Observations, Data Reduction, and Analysis}

The \cxo\ began observing the location of SN~2006gy on 2006 Nov 14.86
using Director's Discretionary Time.  The observation lasted 29.743
ks, and the data were taken with the Advanced CCD Imaging Spectrometer
using an integration time of 3.2~s per frame.  The telescope aimpoint
was on the back-side illuminated S3 chip, and the data were
telemetered to the ground in ``very faint'' mode.  

Data reduction was performed using the CIAO 3.4 software provided by
the \chandra\ X-ray Center\footnote{\url{http://asc.harvard.edu}}.
The data were reprocessed using the CALDB 3.3.0 set of calibration
files (gain maps, quantum efficiency, quantum efficiency uniformity,
effective area) including a new bad-pixel list made with the {\tt
acis\_run\_hotpix} tool.  The reprocessing was done without pixel
randomization that is added during standard processing.  This omission
slightly improves the point-spread function (PSF). The data were
filtered using the standard ASCA grades (0, 2, 3, 4, and 6) 
excluding both bad pixels and software-flagged cosmic-ray events. A
search was done for strong background flaring, but none was found.

Absolute \chandra\ astrometry is typically good to 0\farcs5, and we
sought to tie the \chandra\ frame to the KAIT image to obtain a
reliable identification of the nucleus of NGC 1260 and the SN in the
\chandra\ data.  Several \chandra\ point sources were found using the
CIAO {\tt wavdetect} tool, and their positions were refined using ACIS
Extract version 3.107 (Broos et al.\ 2002).  Three of these sources
had KAIT counterparts, although one had a somewhat poorly determined
\chandra\ position due to its location $\sim$3\arcmin\ off-axis (the
\chandra\ PSF degrades as a function of off-axis angle).  Using all
three sources, we obtained an astrometric correction to the \chandra\
data of 0\farcs329 in right ascension ($\alpha$) and 0\farcs089 in 
declination ($\delta$).  Using the two best counterparts, we obtained 
shifts of $\Delta \alpha = 0\farcs472$ and $\Delta \delta = 0\farcs104$.
We use this latter shift for the rest of our analysis.

Figure~\ref{fig:xrayimage} shows a 0.5--2 keV image of the \chandra\
data after this shift; arrows indicate the KAIT positions of the SN
(red) and galaxy nucleus (blue).  In addition to the raw image,
Figure~\ref{fig:xrayimage} shows a Gaussian-smoothed image and a
maximum-likelihood reconstruction of the data, as well as an image of
the \chandra\ PSF on the same spatial scale.  The maximum-likelihood
reconstruction was made by ACIS Extract using the {\tt
max\_likelihood} procedure available in the IDL Astronomy User's
Library\footnote{\url{http://idlastro.gsfc.nasa.gov/contents.html}};
we went through 200 iterations of the algorithm, using the PSF shown
in the figure.  The PSF was constructed by ACIS Extract through use of
the CIAO tool {\tt mkpsf} based on the off-axis location of the source
and at an energy of 1.49~keV (the \chandra\ PSF is also a function of
energy).  As can be seen, there is excellent agreement between the
locations of the reconstructed sources and the locations of the SN and
host-galaxy nucleus.  This argues strongly that we have, in fact,
detected SN~2006gy and spatially resolved it from the nucleus of
NGC~1260.

We measured counts in the full 0.5--8 keV bandpass from the position
of the SN using a small extraction region to minimize contamination
from the galaxy nucleus.  The extraction region has a radius of
$\sim$0\farcs4, corresponding to about 40\% of the PSF.  Response
files were constructed with the CIAO tools, and ACIS Extract corrected
them for the non-standard extraction region.  The background region is
a source-free annulus centered on the position of the SN with inner
and outer radii of 6\arcsec\ and 14\arcsec, respectively.  Based on
the 241 counts detected in this region, we expect only 0.24 background
counts in our extraction region.  In the restricted energy range of
0.5--2 keV (used for the rest of this paper), we expect only 0.08
background counts in our extraction region.

Four counts were detected in our extraction region, which precludes a
detailed spectral analysis.  However, the counts were all detected
below 2 keV, giving some indication of the spectral shape.  We
assume a thermal plasma spectrum (Raymond-Smith) with $kT = 1$ keV to
estimate the luminosity.  Such thermal spectra have successfully fit
the X-ray spectra of SNe, and temperatures much higher than
this would result in significant emission detectable by \chandra\
(which was not seen).  Based on an assumed reddening toward SN~2006gy
of $E(B-V) = 0.74$ mag, we assume an X-ray absorbing column of $n_H =
4.1 \times 10^{21}~\mathrm{cm}^{-2}$ (Predehl \& Schmidt 1995).  Such
an absorbed thermal plasma observed by \chandra\ would result in a
ratio of 0.5--2 keV to 2--8 keV counts of $\sim$10:1, in accordance
with observations.  We fit this model to the observed 0.5--8 keV
spectrum in Sherpa (Freeman et al.\ 2001) using the statistic of Cash
(1979).  The only free parameter is the overall normalization of the
model.  From the best fit we find an unabsorbed X-ray luminosity
(0.5--2 keV) of $1.65 \times 10^{39}$ erg~s$^{-1}$.

\section{THE DEATH OF A VERY MASSIVE STAR WITH ITS HYDROGEN ENVELOPE INTACT}

\subsection{The Energy Budget and a High-Mass Progenitor}

SN~2006gy has quickly distinguished itself as unique from other SNe in
two important ways.  First, after correcting for distance and
extinction, it is the most luminous SN ever seen, and second, it has
exhibited a remarkably slow rise to its peak luminosity, and has
stayed bright for a very extended time.  SN~2006gy has peaked and is now
on a slow decline, but even after 200~d it is still as luminous as
the peak of a typical SN~Ia.

SN~2006gy was classified as a SN~IIn with narrow hydrogen lines in its
spectrum at early times (Harutyunyan et al.\ 2006), although the
spectrum has notable differences compared with prototypes of this
class.  It dramatically violates the expectation that SNe~II are
generally less luminous than SNe~Ia (Fig.\ 2 includes a fairly typical
Type II SN 1999em), and that SNe~IIn usually take only $\sim$20~d to
reach their peak (Li et al.\ 2003).  SN~2006gy, by contrast, took
$\sim$70~d to gradually climb to its peak. For about 100~d it was more
luminous than $M_R = -21$ mag, brighter than any other SN known to
date.

Simply put, for a supernova to be extremely luminous and to remain
that way for such an extended time is truly spectacular.  Integrating
the light curve in Figure 2 and assuming zero bolometric correction,
we calculate a total radiated energy of $E_{\rm rad} = (1.2 \pm 0.2)
\times 10^{51}$ erg.  This requires either very efficient conversion
of blast-wave kinetic energy into light, or some alternative energy
source.  One or a combination of the three following traditional
mechanisms may power the visual light: (1) H recombination/thermal
radiation of the supernova ejecta, (2) interaction of the supernova
blast wave with the CSM, or (3) energy from radioactive decay of
$^{56}$Ni. Continued observations and probably extensive theoretical
work will be needed to choose decisively between these options, but
here we argue that regardless of which of these three mechanisms is
responsible, the extreme energy budget of SN~2006gy requires that its
very massive progenitor star retained its H envelope until it
exploded.

\begin{figure}
\epsscale{1.05}
\plotone{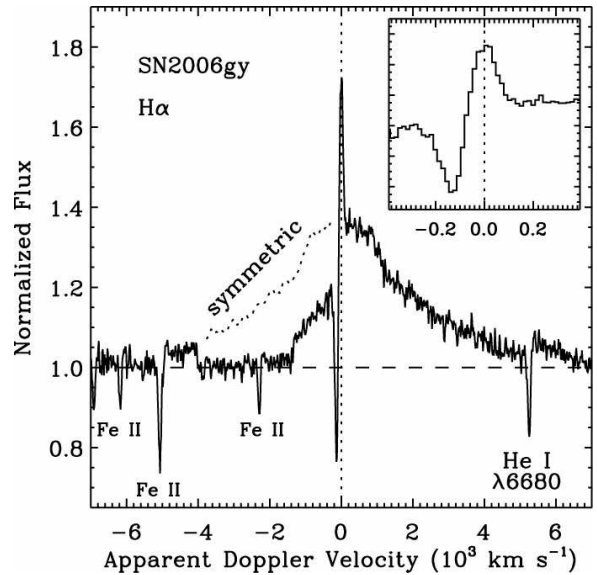}
\caption{The Keck/DEIMOS spectrum of the H$\alpha$ line seen in SN
2006gy, with the flux normalized to the underlying continuum.  The
upper-right inset shows a closer view of the narrow P Cygni line
profile that we believe to be associated with dense unshocked CSM.
The blueshifted narrow absorption trough has a minimum at about $-$130
km s$^{-1}$, reaching $-$260 km s$^{-1}$ at its blue edge.  The other
narrow absorption lines labeled as ``Fe~{\sc ii}'' are Fe~{\sc ii}
$\lambda\lambda$6418, 6433, 6456, 6517.  The dashed line labeled
``symmetric'' is the red side of the broad H$\alpha$ line reflected to
blueshifted velocities, showing what the line shape would be if it
were symmetric. Comparing this to the observed H$\alpha$ profile, we
see significant blueshifted H$\alpha$ absorption from 0 km~s$^{-1}$
out to a sharp blue edge at about $-$4000 km~s$^{-1}$, which we take
to be the dominant speed of the SN blast wave.  At that point, the
blueshifted emission recovers to the level expected for a symmetric
profile, and then gradually declines to the continuum level at about
$-$6000 km~s$^{-1}$, just as on the red side of the line (which
overlaps with He~{\sc i} $\lambda$6680).}
\label{fig:ha}
\end{figure}

\begin{figure*}
\epsscale{0.65}
\plotone{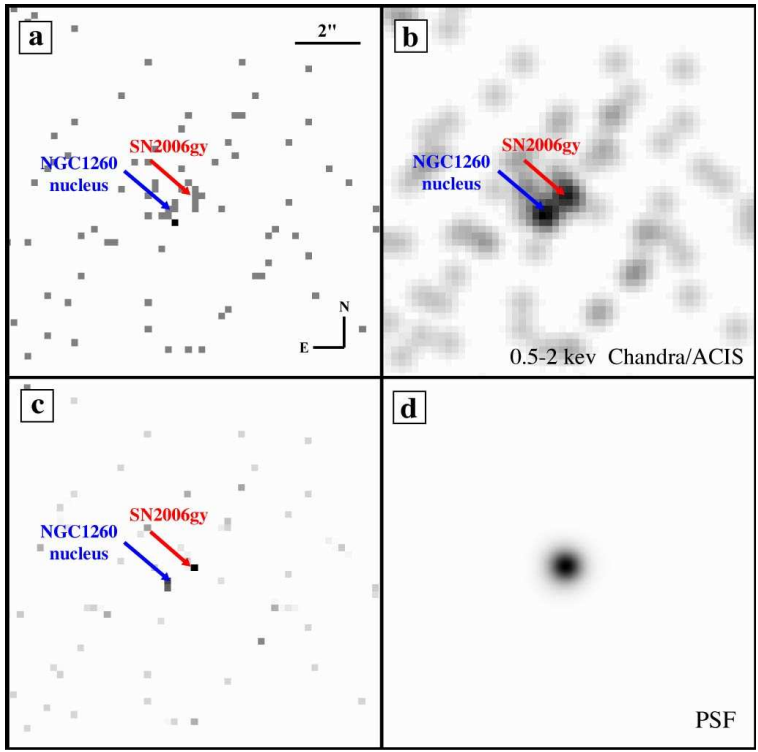}
\caption{Soft-band (0.5--2 keV) \chandra\ images of NGC 1260.  Panel
$a$ shows the raw \chandra\ data (after our astrometric correction)
with red and blue arrows indicating the KAIT positions of the SN and
galaxy nucleus, respectively.  Panel $b$ is a Gaussian-smoothed
version of this image, in which the sources are more clearly apparent.
Panel $c$ is a maximum-likelihood reconstruction of the 0.5--2 keV
image (see text for details).  Panel $d$ shows the \chandra\ PSF at
the location of the galaxy on the same spatial scale as the other
panels.}
\label{fig:xrayimage}
\end{figure*}

The first option of thermal emission from the H-recombination front in
the supernova debris would require a huge ejected mass of order 100
M$_{\odot}$ or more, based simply on the total radiated energy.  A
heavy H envelope might help explain the unusually slow speed of only
about 4000 km~s$^{-1}$ indicated by the H$\alpha$ line (Fig.\
\ref{fig:ha}), and might provide a natural explanation for the long
duration and rise time of the SN because of time needed for energy to
diffuse out of the massive envelope.  Whether or not the SN could
actually radiate efficiently enough to produce the observed luminosity
with this mechanism remains to be proven and should be investigated
with detailed calculations.  For example, at the temperature of the
photosphere defined by the H-recombination front (typically 5000~K to
8000~K), the luminosity of SN~2006gy requires an emitting radius
larger than what we might expect from its observed expansion speed of
4000--4500 km~s$^{-1}$ and age.  Instead of 70~d, the observed peak
luminosity would seem to require an age of 200--380~d since explosion
(assuming linear motion), or rapid deceleration at early times.  Such
rapid deceleration at early times cannot be ruled out by our data.

The second option of powering the visible light entirely with CSM
interaction is problematic, but is difficult to rule out conclusively.
>From the relatively weak soft X-ray flux of SN 2006gy detected by
\chandra, we derive an upper limit to the progenitor star's mass-loss
rate of $\sim 5 \times 10^{-4}$ M$_{\odot}$ yr$^{-1}$ (see \S 3.2).
We find that this falls short of the circumstellar density that would
be needed to power the visual light curve of SN~2006gy by three orders
of magnitude (\S 3.3).  In order to explain the high luminosity in
whole or in part by CSM interaction, one would therefore need to
assume that the X-ray emission is severely quenched and that the {\it
Chandra} detection is erroneous --- but this is difficult to accept,
since we clearly detect soft (unabsorbed, not hard) X-ray emission
from the position of the SN (Fig.\ \ref{fig:xrayimage}).  Even if it
were true, though, a closer look at the demands placed on the
circumstellar density make it difficult to explain with anything other
than a massive star that coincidentally had an LBV outburst just
before the supernova explosion.

Finally, the third option, radioactive decay of $^{56}$Ni, is perhaps
the least problematic, as we will discuss further in \S 3.4.  The main
point of interest is that if this mechanism powers the visual light,
then the high luminosity of SN~2006gy requires a very large Ni mass
that cannot arise from a normal core-collapse SN.  Instead, the large
mass involved would require that SN~2006gy was a pair instability
supernova in which the star's core was obliterated.  If true, {\it SN
2006gy would be the first observed example of a pair-instability
supernova}.  This mechanism also has some potential difficulties, but
they are more along the lines of uncharted theoretical territory,
rather than fundamental physical or observational constraints.
Therefore, SN~2006gy provides fertile ground for important theoretical
work in this area.

\subsection{Limits to the Progenitor's Mass-Loss Rate from X-ray Data}

If we interpret the X-ray emission detected by \chandra\ as the result
of interaction of the outgoing shock with circumstellar material (CSM
interaction), we can place an upper limit on the mass-loss rate of the
progenitor star.  This interaction has been explored in detail (e.g.,
Fransson, Lundqvist, \& Chevalier 1996).  The softness of the X-ray
emission points toward a reverse-shock origin, and we use the
adiabatic case.  A useful form of their eq.\ (3.10) is found in Pooley
et al.\ (2002):

\begin{eqnarray}
 \frac{dL_\mathrm{rev}}{dE} & = 2\times10^{35} \ \zeta(n-3)(n-4)^2 \ 
  T_8^{-0.24}e^{-0.116/T_8} \nonumber\\
 &\times\left(\frac{\dot{M}_{-6}}{V_{w1}}\right)^2
  V_{s4}^{-1}\left(\frac{t}{10~\mathrm{d}}\right)^{-1}~\mathrm{erg\
  s}^{-1}~\mathrm{keV}^{-1},
\end{eqnarray}

\noindent where $\zeta = 0.86$ for solar abundances, $n$ is the index
of the ejecta-density profile [$\rho_\mathrm{SN} \propto t^{-3}
(r/t)^{-n}$], $T_8$ is the temperature in units of $10^8$~K,
$\dot{M}_{-6}$ is the progenitor's steady-state mass-loss rate in
units of $10^{-6}~M_\odot~\mathrm{yr}^{-1}$, $V_{w1}$ is its wind
speed in units of 10~km~s$^{-1}$, $V_{s4}$ is the shock velocity in
units of $10^4$~km~s$^{-1}$, and $t$ is the time since explosion.

The value of $n$ appropriate for SN~2006gy is uncertain, but typical
values for core-collapse SNe are in the range 7--12.  We assume a
temperature of 1~keV, for which $T_8 = 0.116$.  From
Figure~\ref{fig:ha} we take the wind speed to be $\sim$200~km~s$^{-1}$
($V_{w1}=20$), and the shock velocity to be 4500~km~s$^{-1}$
($V_{s4}=0.45$).  The \chandra\ observation took place 87~d after the
explosion.

This implies a mass-loss rate for the progenitor of $1.4 \times
10^{-4}$~M$_{\odot}$~yr$^{-1}$ assuming a steady mass-loss rate of the
progenitor in the decades before explosion, and adopting a SN ejecta
density profile with $n = 12$.  For a profile with $n = 7$, the
mass-loss estimate rises to $5.4 \times 10^{-4}$
M$_{\odot}$~yr$^{-1}$.  This range of mass-loss rates is in good
agreement with observed values in luminous H-rich WN stars (e.g.,
Hamann et al.\ 2006) or quiescent non-outburst LBVs (Smith, Vink, \&
de Koter 2004).  As we discuss below, however, this range of mass-loss
rates falls short of that needed to power the luminosity of SN~2006gy
with CSM interaction by three orders of magnitude.  This is a serious
obstacle to any such model, which must now account for why we observe
a relatively weak and soft (i.e., unabsorbed) X-ray flux from
SN~2006gy.  A likely explanation is that CSM interaction is important
in creating the observed soft X-rays and in causing the emission-line
spectrum of SN~2006gy (especially the broad H$\alpha$ emission), but
that something else drives its visual-wavelength continuum luminosity.
Below, we consider the CSM interaction hypothesis (\S 3.3) as a power
source for SN~2006gy aside from the difficulty posed by X-rays, as
well as an alternative energy source for its radiated luminosity (\S
3.4).

\subsection{A Closer Look at Circumstellar Interaction}

Ofek et al.\ (2007) suggested CSM interaction as a means to power the
visual light of SN~2006gy, but here we wish to make a clear
distinction between two different scenarios.  The first is where the
blast wave from a SN~Ia interacts with dense CSM from a companion star
that provides the hydrogen in the spectrum (the so-called ``Type IIa''
scenario; e.g., Deng et al.\ 2004), as suggested in version 1 (in
astro-ph/0612408) of the recent study by Ofek et al.\ (2007).  This
interpretation had also been suggested previously for the bright SNe
2002ic and 2005gj (Hamuy et al.\ 2003; Deng et al.\ 2004; Aldering et
al.\ 2006).  Note, however, that Benetti et al.\ (2006) have instead
argued in favor of a core-collapse origin for SN~2002ic, so the true
nature of these events is still controversial.  The second type of
scenario would be a blast wave from a core-collapse or
pair-instability supernova from a massive star interacting with its
own ejecta, analogous to the interpretation of the SN~IIn 1994W by
Chugai et al.\ (2004).

We argue here that the first scenario (SN~IIa) is untenable for
SN~2006gy for a number of reasons.  Based in part on a preprint of our
work presented here (version 1 of astro-ph/0612617), Ofek et al.\
(2007) revised their original Type IIa interpretation of SN~2006gy to
include the possibility that it could have been a massive star as we
originally proposed.  The second SN~1994W-like scenario, on the other
hand, is almost certainly relevant to SN~2006gy, but based on the weak
X-ray emission we probably require a different source for the bulk of
the radiated luminosity.  If, for the sake of argument, we demand that
CSM interaction powers the luminosity, we find that the extraordinary
energy demands of SN~2006gy point to a circumstellar environment that
is only likely to be produced by a very massive star that suffered a
rare outburst immediately prior to the SN.  In the case of SN~2006gy,
the luminosity and total energy need to be scaled up by a factor of 40
or more from those for SN~1994W.

In order to power the luminosity of SN~2006gy with CSM interaction,
the environment created by the progenitor star must be extraordinarily
dense.  Ofek et al.\ (2007) originally (version 1 of astro-ph/0612408)
estimated that to achieve the luminosity of SN~2006gy with a shock
plowing into CSM, the progenitor star (or its companion star in a
close binary system) needed to have a wind with an average mass-loss
rate of $\sim$10$^{-2}$ M$_{\odot}$ yr$^{-1}$ in the decades before
explosion.  However, this estimate scales with the adopted wind speed
$V_w$ and inversely with the shock speed $V_s$, which Ofek et al.\
originally took to be $V_w = 10$ km~s$^{-1}$ and $V_s = 10^4$
km~s$^{-1}$.  Instead, though, we observe a much {\it faster} speed of
$V_w \approx 200$ km~s$^{-1}$ in the circumstellar environment
indicated by the narrow P Cygni component in our spectra (Fig.\
\ref{fig:ha}; see \S 3.5), raising this necessary mass-loss rate to
$\sim 0.2$ M$_{\odot}$ yr$^{-1}$ to achieve the same circumstellar
density (Ofek et al.\ 2007 note this in version 2 of their paper,
based on velocities in our Fig.\ \ref{fig:ha}).  We also see a {\it
slower} speed for the SN shock of only $V_s \approx 4000$ km~s$^{-1}$
(Fig.\ \ref{fig:ha}) instead of $10^4$ km~s$^{-1}$ (Ofek et al.\
2007), raising the required progenitor mass-loss rate even further to
about 0.5 M$_{\odot}$ yr$^{-1}$.\footnote{One might suspect that even
this value may underestimate what is required to power SN2006gy.  In a
more detailed analysis of SN~1994W, Chugai et al.\ (2004) required a
similar progenitor mass-loss rate of 0.2 M$_{\odot}$ yr$^{-1}$ for a
short time preceding the SN, yet SN1994W was more than 10 times less
luminous than SN~2006gy.}  Thus, if CSM interaction is to power the
visual light of SN~2006gy, the progenitor was probably an extremely
massive star. Recall, however, that this required value of 0.5
M$_{\odot}$ yr$^{-1}$ is 1000 times above the highest likely value
indicated by X-ray emission, making it problematic (see \S 3.2).  Let
us put this last issue aside for the time being, assuming that the
X-rays are somehow absorbed without hardening the spectrum, so that we
can consider the implications of the CSM interaction hypothesis.

The expansion speed indicated by the H$\alpha$ line (Fig.\
\ref{fig:ha}) is critical for addressing the extent to which
interaction with CSM may power the observed radiation, because the
FWHM$\approx$2400 km s$^{-1}$ of the main intermediate-width
emission component in Figure~\ref{fig:ha} has changed little from the
initial value of FWHM$\approx$2500 km~s$^{-1}$ seen in the H$\beta$
emission feature only a few days after discovery (Harutyunyan et al.\
2006).  (Recall that if the SN is powered by CSM interaction, then the
observed expansion speed traces the blast-wave speed, and not the
decrease in speed expected as the H recombination front progress
deeper into the SN ejecta.)  If the expanding blast wave has only
slowed by about 10\% in the first few months, conservation of momentum
dictates that the mass of swept-up material is only about 10\% of the
ejected mass. Since at least a few M$_{\odot}$ of material needs to be
swept up to power the luminosity of SN~2006gy\footnote{This comes from
the required progenitor mass-loss rate, the duration of the SN at the
time the spectrum in Fig.\ 5 was taken ($t \approx 96$~d), and the
relative speed of the blast wave and circumstellar material: $M =
\dot{M} \times t(V_S/V_w)$, which gives about 2.5 M$_{\odot}$.}, the
mass of the SN ejecta then needs to be {\it at least} 25 M$_{\odot}$.
This clearly rules out a Type Ia event.  Another way to approach the
problem is that if the ejecta only slow by 10\% after discovery, then
only $\sim$20\% of the initial kinetic energy can be converted into
radiation during that time.  The huge radiated energy of SN~2006gy
would then require a SN with $\ga 5 \times 10^{51}$ erg,
again too great a demand for a SN~Ia, even in a double-degenerate
scenario or a super-Chandrasekhar-mass white dwarf.\footnote{Invoking
the hypothesis that the CSM interaction occurred before the first
observation, allowing the observed SN expansion speed to remain
constant, does not help because it cannot account for how the light
curve is powered continually for more than 100~d after that
interaction (the ejecta cool quickly).}  In short, one cannot extract
enough energy from the shock to power the light curve without slowing
down the shock, unless the initial mass and kinetic energy of the SN
ejecta are high.

Even if we somehow allow for very efficient conversion of all the
$10^{51}$ erg of blast-wave kinetic energy into radiation, we must
ask: {\it What type of progenitor star is likely to have had such a
stupendous mass-loss rate?} A rate of 0.5 M$_{\odot}$ yr$^{-1}$ would
be unheard of for a low-mass (2--8 M$_{\odot}$) asymptotic giant
branch (AGB) star, which is the most likely type of star to expect in
the SN~IIa scenario, for which observed mass-loss rates are four to
five orders of magnitude lower (de Jager et al.\ 1988).  Even the most
extreme OH/IR stars have rates below $10^{-4}$ M$_{\odot}$ yr$^{-1}$
(Netzer \& Knapp 1987), while the highest rates during the final and
brief protoplanetary nebula phase reach only (1--2) $\times$ 10$^{-4}$
M$_{\odot}$ yr$^{-1}$ (Bujarrabal et al.\ 2001).  In fact, it is also
more than 4 orders of magnitude larger than the Eddington accretion
rate for a white dwarf, which would be relevant in a common-envelope
scenario.  Even massive stars in their normal (i.e., non-eruptive)
states do not come close to this rate.

The only type of star known to have a mass-loss rate higher than 0.1
M$_{\odot}$ yr$^{-1}$ would be an LBV during a giant eruption (Smith
\& Owocki 2006). Those events typically last about a decade or less
(Van Dyk 2005), which would be of the right order ($t = 150~{\rm d}
\times V_S/V_w$) to account for the required circumstellar environment
of SN 2006gy.  The outbursts are impulsive, so the large masses in
their nebulae (Smith \& Owocki 2006) averaged over the durations of
the visible eruptions yield these mass-loss rates.  If it were the
case that the pre-SN mass-loss event before SN~2006gy was of such
short duration, then we would predict the luminosity of SN~2006gy to
soon plummet rapidly to the late-time luminosity of a normal SN~II.
If such a drop is not observed, it will strengthen the case for the
pair-instability hypothesis discussed next in \S 3.4.  Such a sudden
drop was clearly seen in SN~1994W at roughly day 110 (Chugai et al.\
2004).  

This interpretation, though, forces us back once again to the
hypothesis that the progenitor was an extremely massive star, since
only the most powerful LBV outbursts from the most massive stars with
initial masses above $\sim$100 M$_{\odot}$ are known to have such
high mass-loss rates.  Coincidentally, the mass-loss rate of $\eta$ 
Carinae during its phenomenal 1843 eruption was about 0.5 M$_{\odot}$
yr$^{-1}$ if averaged over 20 years (Smith et al.\ 2003).  Another
such extreme case is SN~1961V in NGC~1058 (Goodrich et al.\ 1989;
Filippenko et al.\ 1995; Van Dyk et al.\ 2002), which is thought to
have had an initial mass well above 100 M$_{\odot}$.  To expect such
an extraordinary feat from a low-mass or intermediate-mass star is
unreasonable even in the most imaginative circumstances.

Further difficulties for the SN~IIa scenario --- and even for
moderately massive progenitors --- arise if we consider geometry.  If
one attempts to account for the unusually dense circumstellar
environment by invoking a high mass-loss rate tidal stripping
``event'' in a close binary or common envelope/merger\footnote{Ignore
for the moment that this hypothetical event needs to be synchronized
with the supernova.}, for example, then this would almost certainly
distributed material in a flattened disk as mass is shed from the
system through the outer Lagrangian point (e.g., Taam \& Ricker 2006).
In that case, however, even with 100\% efficiency in the local
conversion of kinetic energy into radiation, the global fraction of
energy available is only that of the solid angle that can be
intercepted by the disk --- which will probably be less than 10\%.

Altogether, then, there are several clear reasons why the Type IIa
scenario originally advocated by Ofek et al.\ (2007, version 1 of
astro-ph/0612408) fails to power SN~2006gy through CSM interaction.
It is perhaps not surprising, then, that the visual spectrum of
SN~2006gy does not resemble a SN~Ia or the other SN~IIa candidates.
Hamuy et al.\ (2003) argued that SN~2002ic was a variant of the SNe~Ia
phenomenon on the basis of the similarity of its spectral evolution to
that of a diluted version of SN~1991T (Filippenko et al.\ 1992).
While the continuum of the earliest spectrum of SN~2005gj was
relatively featureless, it too developed the prominent broad iron
lines typical of a SN~Ia by two months after explosion (Aldering et
al.\ 2006).  Our earliest spectrum of SN 2006gy is plotted in
Figure~\ref{fig:lick} along with SN~1991T at a similar epoch relative
to explosion.  The only strong spectral feature in the SN~2006gy
spectrum is H$\alpha$.  The weaker features that are present do not
match those of SN 1991T.  In particular, the deep minima in the SN
1991T spectrum near 5700 and 6200~\AA\ are lacking in SN~2006gy.  At
no later epoch did SN~Ia features become visible in SN~2006gy, as can
be seen in the day 71 and 96 spectra plotted in Figures 3 and 4.  We
therefore have no compelling reason to believe that an exploding white
dwarf was present in this event.

We find that conversion of the blast-wave kinetic energy into radiated
luminosity might potentially power SN~2006gy, as has been proposed for
SN~1994W (Chugai et al.\ 2004), but only if the swept-up environment
is consistent with extreme environments observed around the most
massive evolved stars known, such as $\eta$ Carinae.  This agrees with
the conclusions in \S 3.5, where the properties of the circumstellar
nebula independently rule out progenitor stars with initial masses
below 40 M$_{\odot}$; initial masses above 60--80 M$_{\odot}$
are favored.

This last conclusion about the progenitor and its environment should
not be taken lightly.  It requires that an extremely rare event
analogous to the 19th-century eruption of $\eta$ Carinae occurred a
decade or so before the SN explosion.  {\it Why would these two events
be synchronized?}  We are left with a choice: Either this is such an
unlikely event that the underlying power source for SN~2006gy must be
some other mechanism and CSM interaction only contributes a fraction
of the radiated energy (see \S 3.4), or instead, it is an indication
that giant LBV eruptions may be a sign of things to come --- i.e., an
``early warning sign'' of an impending SN.  The second possibility
would be astounding if true, and SN~2006gy may not be alone in this
regard.  SN~1994W (Chugai et al.\ 2004; Sollerman et al.\ 1998),
SN~2001em (Chugai \& Chevalier 2006), and SN~2006jc (Foley et al.\
2007) all show signs of dense environments that were probably produced
by a giant mass-loss event just before the SN.  Smith \& Owocki (2006)
have noted several other cases as well.  SN~2006jc, in particular, was
even observed as a ``supernova imposter'' 2 years before the final 
explosion (Nakano et al.\ 2006; Foley et al.\ 2007; Pastorello et al.\ 
2007). Furthermore, such an outburst preceding the SN event may have 
some theoretical expectation (e.g., Heger \& Woosley et al.\ 2002).  
This may be a profound clue to the fates of the most massive stars.

In any case, it is a marked difficulty for the CSM interaction
hypothesis in general that --- in addition to the softness and
faintness of the detected X-rays noted above --- the light curve,
spectrum, and multiwavelength properties of SN~2006gy differ from
those of other SNe~IIn powered by CSM interaction, such as SNe~1988Z
(Filippenko 1991; Stathakis \& Sadler 1991; Turatto et al.\ 1993),
1995N (Fox et al.\ 2000; Fransson et al.\ 2002), and 1998S (Leonard et
al.\ 2000; Pooley et al.\ 2002).  SN~1988Z was bright in X-ray and
radio emission (Schlegel \& Petre 2006; Van Dyk et al.\ 1993; Williams
et al.\ 2002), unlike SN~2006gy.  The complex and unique spectral
evolution of SN~2006gy will be discussed in a later paper, when more
complete data are available.

\subsection{Initial Thoughts on Radioactive Decay and the
Pair-Instability Hypothesis for SN~2006gy}

In previous sections, we have noted some obstacles, primarily
observational in nature, with simple fireball or CSM interaction
models as the engine for SN~2006gy.  Although a suitable choice of
extreme conditions may allow them to work, at least in part, our
observation of soft unabsorbed X-rays from SN~2006gy and the
corresponding upper limits to the progenitor star's mass-loss rate
make it worthwhile to consider other options.  Powering SN~2006gy with
radioactive decay does not suffer from these problems, because this
mechanism is known to work in other SNe.  The question here centers
around whether it is plausible to simply scale up the $^{56}$Ni
decay that powers fainter SNe, how that large mass of Ni may be
created, and what happens to the radiation mechanisms in that extreme
case.  If SN~2006gy is powered by radioactive decay, the large Ni mass
would require a pair-instability SN, as discussed below.

Scannapieco et al.\ (2005) presented model light curves for
pair-instability SNe, where the progenitor stars were assumed to be
red supergiants.  The resulting light curves showed an initial small
peak, but then a long, slow rise to maximum powered by $^{56}$Ni and
$^{56}$Co decay.  Some of their models get nearly the peak luminosity of
SN~2006gy, but they rise more slowly to maximum than SN~2006gy did.
However, their calculations were for zero metallicity, non-rotating
stars with no pre-SN mass loss.  Different assumptions about the
metallicity, mass-loss, and the presence of rotational mixing may
change things considerably (e.g., Maeder 1987; Yoon \& Langer 2005; 
Woosley \& Heger 2006). Also, if the progenitor of SN~2006gy had a
small radius as we expect for an LBV (RSGs are not observed at high
luminosity in normal-metallicity stars), then the initial peak may be
lost due to adiabatic cooling, and the delayed rise after $\sim$50~d 
would be dominated by $^{56}$Co decay.  Interestingly, this is
similar to the case of SN~1987A, where the progenitor was a blue
supergiant with a small radius, and where its late (70--100~d) peak
was powered by radioactive $^{56}$Co decay.  SN~2006gy took a
similarly long time to reach its peak luminosity, and its light
curve thus far has a shape resembling that of SN~1987A (Fig.\ 2), 
but it was 250 times more luminous. 
 
In addition, the pair-instability models of
Scannapieco et al.\ (2005) predict slow 
expansion speeds of $\sim$5000 km~s$^{-1}$ and the
presence of H in the spectrum, again compatible with
SN~2006gy.  These clues are tantalizing, and it would be interesting
to see models for pair-instability SNe at metallicity closer to solar
values and with compact progenitors.  This is still somewhat virgin
territory and will require continued observational constraints and
detailed calculations to find a suitable model that will work for the
case of SN~2006gy.  Below we sketch a plausibility argument for the
hypothesis that SN~2006gy was a pair-instability SN based simply on
the required power source for its radiated luminosity.

The $R$-band magnitude at the peak of SN~2006gy was at least as bright as
$-$21.8, but could have been significantly brighter because of our
conservative assumptions for the reddening, as noted in \S\S 2.1 and
2.2.  Assuming no bolometric correction (again, conservative), this
corresponds to a peak luminosity of
$\ga (1.7 \pm 0.3) \times 10^{44}$ erg s$^{-1}$.  If this peak
luminosity traces the instantaneous decay rate (Arnett 1982), we can
estimate the necessary mass of initial nickel in the $^{56}$Ni
$\rightarrow$ $^{56}$Co $\rightarrow$ $^{56}$Fe decay.  With a late
peak at $t \approx 70$~d, this will put us well into cobalt decay
instead of nickel, as noted above.  The radiated luminosity from
cobalt decay (Sutherland \& Wheeler 1982) is

\begin{eqnarray}
L &= &1.42\times10^{43}~{\rm erg}~{\rm s}^{-1} e^{-t/111~{\rm d}} \ M_{\rm
    Ni}/{\rm M}_{\odot} \\
  &= &8\times10^{42}~{\rm erg}~{\rm s}^{-1} \ M_{\rm Ni}/{\rm M}_{\odot}, \nonumber
\end{eqnarray}

\noindent
where $M_{\rm Ni}$ is the
initial $^{56}$Ni mass.  The extreme luminosity of SN~2006gy, then,
would require an extraordinarily high Ni mass of roughly 22
M$_{\odot}$ to be synthesized in the explosion.  This can be scaled
down somewhat if CSM interaction contributes part of the energy, but
unless that interaction dominates the light output, this large Ni mass
cannot be explained with a core-collapse SN.  (Compare this to a
normal SN~II arising from a star of 15--20 M$_{\odot}$,
with a typical Ni mass of about 0.07 M$_{\odot}$.)

The large Ni mass implicates a progenitor star that began its life
with a mass well above 100 M$_{\odot}$.  The consequences of this are
potentially far-reaching, and could turn out to be the most
interesting result of this study: namely, the only way to get such an
extraordinarily high Ni mass to power the radiated energy would be
from a pair-instability supernova, where the star's core is
obliterated instead of collapsing to a black hole (Barkat et al.\
1967; Fraley 1968; Bond et al.\ 1984; Heger \& Woosley 2002).  This
type of supernova is only expected to occur in extremely massive
stars.  For the mechanism to work in the modern universe, even the
most massive stars would need to retain most of their initial massive
envelopes, providing a self-consistent interpretation of SN~2006gy in
light of other evidence for its high mass discussed here. This is not
wild speculation --- it may even be the most promising explanation ---
but it deserves close scrutiny because of its far-reaching importance.

As SN~2006gy continues to evolve, it will become easier to determine
if $^{56}$Co decay or CSM interaction is the power source.  If CSM
interaction drives the visible light, we might expect the light curve
to plummet precipitously, down to the luminosity of a normal SN~II,
when the shock reaches the outer extent of the LBV shell.  Such a drop
occurred in SN~1994W, although the light-curve shape of SN~2006gy so
far is quite different from that of SN~1994W (Fig.\ 2).  On the other
hand, if SN~2006gy continues to decay smoothly from its peak, like
SN~1987A but at an elevated luminosity, then it was almost certainly a
pair-instability SN event because of the large nickel mass required.
So far, SN~2006gy shows no sign of plummeting --- in fact, the latest
photometry seems to imply that it is settling onto a plateau.

Of course, SN~2006gy could be a combination of both CSM interaction
and pair instability. Any very massive star capable of suffering a
pair-instability SN is likely to have a strong stellar wind in its
late pre-explosion stages anyway, consistent with the values of (1--5)
$\times 10^{-4}$ that we infer from the X-ray interaction.
The pair-instability SN models of Heger \& Woosley (2002) predict
mass-loss pulses that precede the final explosion.  In fact, the
observed optical spectrum of SN~2006gy {\it requires} that CSM
interaction is occurring at some level, but the critical question is
whether this interaction is capable of powering the enormous continuum
luminosity of SN~2006gy. Current indications are that it cannot.

\begin{figure*}
\epsscale{0.98}
\plotone{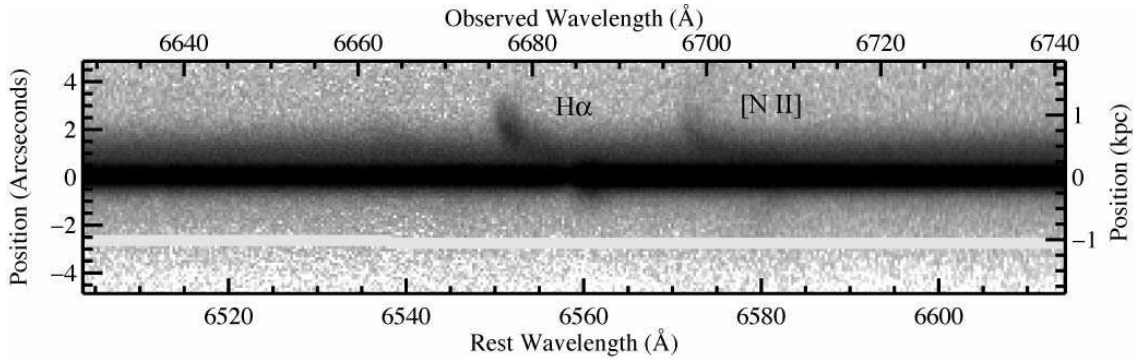}
\caption{The long-slit Keck/DEIMOS spectrum of SN~2006gy and NGC~1260
  in the region around H$\alpha$.  It includes the central point
  source SN~2006gy at the zero-offset position, plus extended emission
  from the host galaxy NGC~1260 on either side of it.  The extended
  H$\alpha$ and [N~{\sc ii}] emission, which follows the rotation
  curve of the galaxy and has an [N~{\sc ii}]/H$\alpha$ intensity ratio
  typical of H~II regions, indicates that NGC~1260 does have active star
  formation. The light row below SN~2006gy is a bad row in the CCD and
  has been masked. }
\label{fig:longslit}
\end{figure*}

\subsection{A Massive Circumstellar LBV Nebula}

Independent of the energy-budget arguments, the properties of the
unshocked circumstellar gas around the progenitor of SN~2006gy are
also consistent with the interpretation that it was a very massive
star, and provide critical clues that strongly refute the hypothesis
that it was powered by the Type Ia explosion of a low-mass star
interacting with dense CSM.  The high-resolution
spectrum in Figure~\ref{fig:ha} contains a narrow component to the
H$\alpha$ line, which also exhibits a clear P Cygni absorption
profile.  It indicates that the SN is expanding into a hydrogen-rich
dense stellar wind or outflowing circumstellar nebula of the progenitor
star, which has an expansion speed of 130--260 km s$^{-1}$ indicated
by the absorption component.  This same narrow absorption component is
seen in other lines in the spectrum of SN~2006gy, such as He~{\sc i}
(He~{\sc i} $\lambda$6680 and Fe~{\sc ii} lines are shown in Fig.\ 4),
Si~{\sc ii}, Fe~{\sc ii}, Ca~{\sc ii}, O~{\sc i}, etc.  The narrow
He~{\sc i} lines are unusual, and may suggest He-enriched material in
the CSM.

This expansion speed is a critical clue to the nature of the
progenitor star that cannot be neglected.  It is much faster than
typical wind speeds of AGB stars (10--20 km~s$^{-1}$), effectively
ruling out the interpretation of SN~2006gy as a SN~IIa.  While it is
unclear if the expansion speed itself is in direct conflict with an
interpretation involving a common envelope mass-loss phase (e.g., Taam
\& Ricker 2006, and references therein), as suggested by Livio \&
Riess (2003) to explain the properties of SN 2002ic, and in the first
version of Ofek et al.\ (2007) to explain SN 2006gy, that
interpretation is ruled out for SN~2006gy based on the energy budget
(see \S 3.3).  This speed is also too fast for a RSG wind (20--40 km
s$^{-1}$), making it difficult to believe that the progenitor star had
an initial mass in the range 10--40 M$_{\odot}$.  Moreover, the speed
is an order of magnitude too slow for the wind of an O-type
supergiant, H-rich WN, or Wolf-Rayet (WR) star progenitor.  On the
other hand, this speed is entirely consistent with an LBV wind or
nebula (e.g., Smith et al.\ 2004; Smith 2006).  Similar absorption
speeds were seen in the narrow P Cygni absorption of SN~1998S, which
Fassia et al.\ (2001) also interpreted as a prior blue-supergiant
phase.  Chugai et al.\ (2002), however, interpreted it somewhat
differently as a fast blue-supergiant wind sweeping into a
red-supergiant wind.  Significant acceleration of the slow
red-supergiant wind would require a swept-up mass comparable to the
fast-wind mass shortly before the SN, which makes this scenario
implausible in the case of SN~2006gy because of the large mass
implied.  The typical LBV ejecta speed agrees well with our
constraints from \S 3.3.

Narrow blueshifted absorption components similar to H$\alpha$ are seen
in a number of other lines throughout the spectrum of SN~2006gy along
with some relatively broad blueshifted absorption (Figs.~\ref{fig:lick}
and \ref{fig:ha}).  Those absorption features are not always present
(Fig.~\ref{fig:lick}), while narrow H$\alpha$ emission remains.  Thus,
we cannot be certain that the narrow emission and absorption
components of H$\alpha$ constitute a true P Cygni scattering profile,
so we consider both cases here.  For each case, the luminosity
of the narrow H$\alpha$ emission component is a relevant quantity.  At
a distance of 73 Mpc, the luminosity of the narrow emission component
of the H$\alpha$ line on day 96 (Fig.~\ref{fig:ha}) is $L_{{\rm H}\alpha}
\approx (1.3 \pm 0.3) \times 10^6$~L$_{\odot}$ (the absolute flux was
calibrated by scaling the red continuum to match observed KAIT
photometry at the appropriate date and correcting for $A_R \approx 1$
mag).  Note that the true luminosity may be somewhat larger than this
because the apparent luminosity may be reduced by the blueshifted
narrow absorption.

If the narrow H$\alpha$ component arises in an unshocked CSM wind, we
can make a rough estimae of the density immediately outside the radius
of the shock, given by $R_{\rm s} = V_{\rm s} t$, where we again take
$V_{\rm s}$=4000~km~s$^{-1}$ and $t = 96$~d is the time the Keck
spectrum was taken.  Such estimates are plagued with uncertainties in
the ionization fraction and H mass fraction, so the estimate below is
a lower limit assuming fully ionized pure H gas.  Following equation
(1) of Chugai \& Danziger (2003), for example, our measured value of
$L_{{\rm H}\alpha}$ implies a density of roughly 2$\times$10$^8$ cm$^{-3}$
just outside $R_{\rm s} \approx 3.3 \times 10^{15}$~cm (adopting
$\alpha^{\rm eff}_{{\rm H}\alpha} = 8.64 \times 10^{-14}$ cm$^3$
s$^{-1}$ for the Case~B H$\alpha$ recombination coefficient as noted
below).  If $V_{\rm w}$ is taken to be 200 km s$^{-1}$, this implies a
mass-loss rate for the progenitor star of roughly 0.01--0.02
M$_{\odot}$ yr$^{-1}$.  While this is an exceptionally high mass-loss
rate, higher than what we infer from the X-ray emission (\S 3.2), it
still falls short of what is required to power the visual luminosity
of SN~2006gy by more than a factor of 10--20.  It is interesting,
however, that this value is comparable to progenitor mass-loss rates
estimated for other SNe~IIn with similar narrow H$\alpha$ P Cygni
features from the unshocked CSM, such as SNe 1997ab and 1997eg
(Salamanca et al.\ 1998, 2002).

If the narrow H$\alpha$ emission component arises instead from
unshocked ionized gas in a detached CS shell nebula, however, then
the mass implied would add yet another requirement that the progenitor
star was very massive.  It may arise in a circumstellar shell like the
Homunculus Nebula of $\eta$~Carinae (Smith 2006), for example.  Using
$L_{{\rm H}\alpha}$, and assuming that the line originates from a
circumstellar shell nebula of constant density, the ionized gas mass
can be expressed as

\begin{equation}
M_{{\rm H}\alpha} \approx \frac{m_{\rm H} L_{{\rm H}\alpha}}{h\nu 
\alpha^{\rm eff}_{{\rm H}\alpha} n_e}, 
\end{equation}

\noindent 
where h$\nu$ is the energy of an H$\alpha$ photon,
$\alpha^{\rm eff}_{{\rm H}\alpha} = 
8.64 \times 10^{-14}$ cm$^3$ s$^{-1}$ is the
Case~B H$\alpha$ recombination coefficient, and $n_e$ is the average
electron density.  This yields $M_{H\alpha} \approx 11.4$~M$_{\odot}$
($L_{{\rm H}\alpha}/n_e$).  We do not know the electron density in the
nebula around SN~2006gy, but values of 10$^5$--10$^6$ cm$^{-3}$ are
the highest densities typically seen in young LBV nebulae like the one
around $\eta$~Carinae (Smith 2006).  With the observed H$\alpha$
luminosity and densities of this order, the nebular mass is probably
above 5~M$_{\odot}$, and it could plausibly be as high as 20--30
M$_{\odot}$.  Lower densities typically seen in circumstellar nebulae
around lower-mass stars would require implausibly high emitting masses
to account for the observed radiation, exceeding their own stellar
masses.  Environments this massive obviously cannot be produced by
low-mass stars and are not seen around moderately massive stars of
20--40 M$_{\odot}$, but they are quite typical of the nebular shells
around LBVs with $L > 10^6$ L$_{\odot}$ (Smith \& Owocki 2006), which
descend from stars with initial masses of 80--150 M$_{\odot}$.  Such
large masses are consistent with the $\ga$12.5 M$_{\odot}$ nebula
around $\eta$~Car (Smith et al.\ 2003).

Thus, the flux of the narrow H$\alpha$ component that we observe is
only likely to arise in the circumstellar nebula of an extremely
massive star.  Taken together, this high mass and the shell's
expansion speed give self-consistent evidence that the progenitor star
was indeed very massive.  This line of reasoning is independent of the
uncertainty associated with the mechanism that powers the radiated
energy of the SN.  It is also consistent with the presence of strong 
hydrogen lines in the spectrum, since LBVs have not yet shed their H 
envelopes. Although dominated by hydrogen, LBV shells also have elevated 
helium abundances, consistent with the presence of narrow He~{\sc i} lines 
in the spectrum of SN~2006gy.  If SN~2006gy really is surrounded by a
dense LBV nebula like that of $\eta$ Carinae, then we might expect to
see strong, narrow emission lines of [N~{\sc ii}]
$\lambda\lambda$6548, 6583 in its late-time spectral evolution, since
LBV nebulae like that of $\eta$ Car tend to be enriched with CNO-cycle
ashes (Smith \& Morse 2004).

\subsection{Do We Expect Massive Stars in the Host Galaxy?}

SN~2006gy has been compared (Ofek et al.) to two peculiar supernovae,
SN~2002ic and SN~2005gj (Fig.\ 2), which have been proposed as SNe~Ia
interacting with dense CSM (the so-called ``Type IIa'' SNe) as noted
earlier.  One factor that motivated Ofek et al.\ (2007) to originally
favor the SN~IIa hypothesis was that the host galaxy, NGC 1260, was
apparently not a star-forming galaxy.  It should not have massive
stars, because S0 galaxies are dominated by old stellar populations.

We note, however, that the SN host is actually a peculiar S0/Sa
galaxy with infrared (IR) emission from dust.  NGC~1260 was detected by the
{\it Infrared Astronomical Satellite} ({\it IRAS}), and Meusinger,
Bruzendorf, \& Krieg (2000) give an infrared luminosity of
log($L_{IR}/{\rm L}_{\odot}) = 9.85$.  According to Kennicutt (1998), 
this would translate to a star-formation rate of $\sim$1.2 M$_{\odot}$
yr$^{-1}$, which is certainly high enough to permit this galaxy to
host some massive young stars.  Furthermore, we detect
extended H$\alpha$ and [N~{\sc ii}] $\lambda$6583 emission from the
galaxy in our spectra; Figure~\ref{fig:longslit} presents the original
long-slit Keck spectrum before the H$\alpha$ profile of SN~2006gy was
extracted, revealing extended emission from gas that follows the
rotation curve of the host galaxy.  These emission lines, having
intensity ratios typical of H~II regions, are indicative of current 
star formation and are absent in non-star-forming galaxies.

A related point concerns the statistics involved.  SN~2006gy is the
most luminous SN seen to date, but it is also spectrally peculiar,
almost in a class by itself.  Its unusual nature would not be at all
surprising, in principle, if its origin were the explosion of a $>$100
M$_{\odot}$ star, since these stars are so phenomenally rare to begin
with.  On the other hand, if it results from normal evolution for low-mass
stars or even moderately massive stars of 10--40 M$_{\odot}$, then we would
expect such events to be more common.

\section{SUMMARY: EXPLOSION AS A MASSIVE LBV AND THE RELEVANCE OF A
PAIR-INSTABILITY SUPERNOVA}

All available observations are broadly consistent with the hypothesis
that the progenitor of SN~2006gy was a very massive star that retained
a massive hydrogen envelope until it exploded.  Retaining this
envelope does not mean that the progenitor was a RSG; the most
luminous stars evolve to the LBV phase before losing their envelopes,
and during that phase they are hot supergiants with relatively small
radii.  This can strongly affect the early light-curve shape.  A mass
below 60 M$_{\odot}$ may be possible if the event was powered by CSM
interaction, but then one must invoke exceptional conditions
inconsistent with observed properties of stars below that mass.  If
CSM interaction dominates, we find it more likely that the progenitor
star had an initial mass of 100--150 M$_{\odot}$, although we still
lack a satisfactory explanation for the weak unabsorbed X-rays in that
case.

By contrast, the huge radiated luminosity, the long duration, the
presence of hydrogen in the spectrum, the low expansion speed of the
SN ejecta, and the various critical clues from the circumstellar
environment are all consistent with the hypothesis that this event was
powered by a pair-instability supernova that also has some moderate
CSM interaction, implying that the progenitor star's initial mass may
have been near the upper mass limit for stars of $\sim$150 M$_{\odot}$
(Figer 2005).  Regardless of the power source, several clues hint that
the progenitor star may have resembled the LBV star $\eta$ Carinae.

If this hypothesis of explosion as a massive LBV is correct, it would
have important consequences for our understanding of stellar
evolution.  It is currently thought that variability in the LBV phase
is responsible for the mass shedding that marks the transition from
the end of core H burning to core He burning, after which a star
appears as a He-rich WR star (Abbot \& Conti 1987; Langer et al.\
1994; Smith \& Owocki 2006; Smith et al.\ 2004).  During this brief
evolutionary phase, a massive star might undergo sequential bursts of
mass loss when it can repeatedly shed more than 10 M$_{\odot}$ of
material in a decade (Smith \& Owocki 2006).  These events are seen in
other galaxies as faint SNe~IIn, or ``supernova impostors'' (Van Dyk
2005, and references therein). They may dominate the mass loss of the
most massive stars, shedding more total mass than line-driven winds
during the star's lifetime (Smith \& Owocki 2006).  Consequently, LBV
stars are frequently surrounded by circumstellar nebulae with masses
of order 10 M$_{\odot}$, like the one that may reside around
SN~2006gy.  It would appear that one of these events may have occurred
within a decade or so immediately preceding SN~2006gy.

The core He burning WR phase that should follow after the massive
hydrogen envelope is stripped away is expected to last a few hundred
thousand years before the star reaches even more advanced stages of
nuclear burning and finally explodes (Abbott \& Conti 1987).  If LBVs
explode before reaching the WR phase, though, it means that they could
be in more advanced stages of nuclear burning than currently predicted
by stellar evolution theory. SN~2006gy adds to mounting evidence
(e.g., Smith \& Owocki 2006; Kotak \& Vink 2006; Gal-Yam et al.\ 2007;
Smith 2007) that stars may explode ``early'' during the LBV phase, and
it hints that reaching the pair instability could be a reason for
this.  

It seems intuitively possible, although difficult to prove, that it
would be the most massive LBVs above $\sim$100 M$_{\odot}$ that are
more likely to explode prematurely as they have a greater burden of
removing their massive envelopes before transitioning to WR stars.
Gal-Yam et al.\ (2007) have drawn a similar conjecture, considering
LBVs as the most likely progenitors of SNe IIn.  If the most massive
stars can indeed explode before the WR phase, then our current
ignorance of the instability underlying the LBV phase presents a
critical challenge.  The possibility that SN 2006gy could have been a
pair-instability supernova weighs heavily upon the importance of
understanding these LBVs as well. SN~2006gy may be giving us a clue
that the wild instability of the most luminous LBVs like
$\eta$~Carinae could be early warning signs of a massive star's
imminent demise, and there may be theoretical reasons to think this is
the case.  One implication is that we had better keep a watchful eye
on $\eta$~Carinae.

The chief reason why pair-instability SNe are expected to occur for
high-mass stars in the early universe is because their low metal
content is expected to reduce their mass-loss rates, causing them to
retain their massive H envelopes (Heger et al.\ 2003; Heger \& Woosley
2002; although see Smith \& Owocki 2006).  Also, the initial mass
function of the first stars is thought to have been skewed to higher
masses due to the lack of metal cooling and consequent fragmentation
in the star-formation process (e.g., Bromm \& Larson 2004). SN~2006gy
may have been a very massive star that exploded as an LBV {\it before}
it could shed its H envelope, and it may have done so by the
pair-instability mechanism. 

The fact that SN~2006gy was able to
explode successfully instead of winking away into a black hole has
far-reaching implications. In particular, one primary goal of the {\it
James Webb Space Telescope} will be to search for these first
explosions in the universe, and the brilliant display of SN~2006gy may
bode well for the possibility of their infrared detection at high
redshift.

\acknowledgments

\small

This study is based in part on data obtained at the W. M.\ Keck
Observatory, made possible by the generous financial support of the
W. M.\ Keck Foundation. KAIT was made possible by donations from Sun
Microsystems, Inc., the Hewlett-Packard Company, AutoScope
Corporation, Lick Observatory, the National Science Foundation, the
University of California, the Sylvia \& Jim Katzman Foundation,
and the TABASGO Foundation. A.V.F.'s supernova group at U.C. Berkeley 
is supported by NSF grant AST--0607485 and by the TABASGO Foundation, 
while J.C.W.\ and R.Q.\ are supported by NSF grant AST--0406740.  A.V.F.\ 
and J.S.B.\ are partially supported by a grant from the Department of
Energy (DE--FC02--06ER41453).  D.P.\ gratefully acknowledges the support
provided by NASA through \chandra\ Postdoctoral Fellowship grant
PF4--50035 awarded by the \chandra\ X-ray Center, which is operated by
the Smithsonian Astrophysical Observatory for NASA under contract
NAS8--03060.  We thank James Graham and Marshall Perrin for assistance
with the Lick AO observations and data reduction; their work has been
supported by the NSF Science and Technology Center for Adaptive
Optics, managed by the University of California at Santa Cruz under
cooperative agreement No. AST--9876783.  We thank the DEEP team, and
especially Michael C. Cooper, for their hard work and assistance with
the DEIMOS reduction pipeline.



\begin{references}

\reference{} Abbot, D. C., \& Conti, P. S.\ 1987, ARAA, 25, 113

\reference{} Akerlof, C. W., et al.\ 2003, PASP, 115, 132

\reference{} Aldering, G., et al.\ 2006, ApJ, 650, 510

\reference{} Arnett, W. D.\ 1982, ApJ, 253, 785

\reference{} Baraffe, I., Heger, A., \& Woosley, S. E.\ 2001, ApJ, 550, 890

\reference{} Barkat, Z., Rakavy, G., \& Sack, N.\ 1967, Physical
Review Letters, 18, 379

\reference{} Benetti, S., Cappellaro, E., Turatto, M., Taubenberger,
S., Harutyunyan, A., \& Valenti, S.\ 2006, ApJ, 654, L129

\reference{} Bessell, M. S.\ 1999, PASP, 111, 1426

\reference{} Bond, J. R., Arnett, W. D., \& Carr, B. J.\ 1984, ApJ, 280, 825

\reference{} Bouret, J. C., Lanz, T., \& Hillier, D. J.\ 2005, A\&A,
438, 301

\reference{} Bromm, V., \& Larson, R. B.\ 2004, ARAA, 42, 79

\reference{} Broos, P.~S., Townsley, L.~K., Getman, K., \& Bauer,
F.~E.\ 2002, ACIS Extract, An ACIS Point Source Extraction Package
(University Park: Pennsylvania State Univ.)

\reference{} Bujarrabal, V., Castro-Carrizo, A., Alcolea, J., \&
Sanchez-Contreras, C.\ 2001, A\&A, 377, 868

\reference{} Cardelli, J. A., Clayton, G. C., \& Mathis, J. S.\ 1989,
ApJ, 345, 245

\reference{} Cash, W.\ 1979, \apj, 228, 939

\reference{} Chevalier, R.~A., Fransson, C., \& Nymark, T.~K.\ 2006,
ApJ, 641, 1029

\reference{} Chugai, N. N., \& Chevalier, R.~A.\ 2006, ApJ, 641, 1051

\reference{} Chugai, N. N., \& Danziger, I.J.\ 2003, Ast.\ Letters,
29, 649

\reference{} Chugai, N. N., Blinnikov, S. I., Fassia, A., Lundqvist, P.,
Meike, W. P. S., \& Sorokin, E. I.\ 2002, MNRAS, 330, 473

\reference{} Chugai, N. N., Blinnikov, S. I., Cumming, R. J., Lundqvist,
P., Bragaglia, A., Filippenko, A. V., Leonard, D. C., Matheson, T., \&
Sollerman, J.\ 2004, MNRAS, 352, 1213

\reference{} Deng, J., Kawabata, K. S., Ohyama, Y., Nomoto, K.,
Mazzali, P. A., Wang, L., Jeffery, D. J., Iye, M., Tomita, H., \&
Yoshii, Y.\ 2004, ApJ, 605, L37

\reference{} Faber, S. M., et al.\ 2003, \procspie, 4841, 1657

\reference{} Fassia, A., et al.\ 2001, MNRAS, 325, 907

\reference{} Figer, D. F.\ 2005, Nature, 434, 192

\reference{} Filippenko, A. V. 1997, ARAA, 35, 309

\reference{} Filippenko, A. V. 1982, PASP, 94, 715

\reference{} Filippenko, A. V. 1991, in SN 1987A and Other Supernovae,
     ed. I. J. Danziger and K. Kj\"ar (Garching: ESO), 343

\reference{} Filippenko, A. V. 2003, in From Twilight to Highlight: 
   The Physics of Supernovae, ed. W. Hillebrandt \& B.
   Leibundgut (Berlin: Springer-Verlag), 171

\reference{} Filippenko, A. V., Barth, A. J., Bower, G. C., Ho, L. C.,
Stringfellow, G. S., Goodrich, R. W., \& Porter, A. C.\ 1995, AJ, 110, 2261

\reference{} Filippenko, A. V., et al.\ 1992, ApJ, 384, L15

\reference{} Foley, R. J., Li, W., Moore, M., Wong, D. S., Pooley, D., \&
Filippenko, A. V.\ 2006, CBET, 695, 1

\reference{} Foley, R.~J., Smith, N., Ganeshalingam, M., Li, W.,
Chornock, R., \& Filippenko, A.~V.\ 2007, ApJ, 657, L105

\reference{} Foley, R. J., et al.\ 2003, \pasp, 115, 1220

\reference{} Fox, D., et al.\ 2000, MNRAS, 319, 1154

\reference{} Fraley, G.~S.\ 1968, \apss, 2, 96

\reference{} Fransson, C., Lundqvist, P., \& Chevalier, R. A.\ 1996,
ApJ, 461, 993

\reference{} Fransson, C., et al.\ 2002, ApJ, 572, 350

\reference{} Freeman, P., Doe, S., \& Siemiginowska, A.\ 2001, Proc.\
SPIE, 4477, 76

\reference{} Fullerton, A. W., Massa, D. L., \& Prinja, R. K.\ 2006, ApJ,
637, 1025

\reference{} Galama, T. J., et al.\ 1998, Nature, 395, 670

\reference{} Gal-Yam, A., et al.\ 2007, ApJ, 656, 372

\reference{} Goodrich, R. W., Stringfellow, G. S., Penrod, G. D., \&
Filippenko, A. V.\ 1989, ApJ, 342, 908

\reference{} Hamann, W.R., Gr\"{a}fner, G., \& Lierman, A.\ 2006,
A\&A, 457, 1015

\reference{} Hamuy, M., \&  Suntzeff, N. B.\ 1990, AJ, 99, 1146

\reference{} Hamuy, M., et al.\ 2003, Nature, 424, 651

\reference{} Harutyunyan, A., Benetti, S., Turatto, M., Cappellaro,
E., Elias-Rosa, N., \& Andreuzzi, G..\ 2006, CBET, 647, 1

\reference{} Heger, A., \& Woosley, S. E.\ 2002, ApJ, 567, 532

\reference{} Heger, A., Fryer, C. L., Woosley, S. E., Langer, N., \&
Hartmann, D.H.\ 2003, ApJ, 591, 288

\reference{} Humphreys, R.M., \& Davidson, K.\ 1979, ApJ, 232, 409

\reference{} de Jager, C., Nieuwenhuijen, H., \& van der Hucht, K. A.\
1988, A\&AS, 72, 259

\reference{} Kennicutt, R. C. 1998, ARAA, 36, 189

\reference{} Kotak, R., \& Vink, J. S.\ 2006, A\&A, 460, L5

\reference{} Langer, N., Hamann, W. R., Lennon, M., Najarro, F.,
Pauldrach, A. W. A., \& Puls, J., 1994, A\&A, 290, 819

\reference{} Leonard, D. C., et al.\ 2000, ApJ, 536, 239

\reference{} Leonard, D. C., et al.\ 2002, PASP, 114, 35

\reference{}Li, W., Filippenko, A. V., Chornock, R., \& Jha, S.\ 2003a,
\apj, 586, L9

\reference{}Livio, M., \& Riess, A. G. 2003, ApJ, 59, L93

\reference{} Li, W., Filippenko, A. V., Van Dyk, S. D., Hu, J., Qiu, Y.,
Modjaz, M., \& Leonard, D. C.\ 2003b, PASP, 114, 403

\reference{} Lloyd, J. P., Liu, M. C., Macintosh, B. A., Severson, S. A.,
Deich, W. T., \& Graham, J. R.\ 2000, SPIE Proceedings, 4008, 814

\reference{} Maeder, A.\ 1987, A\&A, 178, 159

\reference{} Matheson, T., Filippenko, A. V., Ho, L. C., Barth, A. J.,
  \& Leonard, D. C.\  2000, AJ, 120, 1499

\reference{} Max, C. E., et al.\ 1997, Science, 277, 1649

\reference{} Meusinger, H., Bruzendorf, J., \& Krieg, R.\ 2000, A\&A,
363, 933

\reference{} Miller, J. S., \& Stone, R. P. S.\ 1993, Lick
  Obs. Tech. Rep. 66 (Santa Cruz: Lick Obs.)

\reference{} Nakano, S., Itagaki, K., Puckett, T., \& Gorelli, R.\
2006, CBET, 666, 1

\reference{} Netzer, N., \& Knapp, G. R.\ 1987, ApJ, 323, 734

\reference{} Ofek, E. O., et al.\ 2007, ApJ, 659, L13

\reference{} Pastorello, A., et al.\ 2007, Nature, in press

\reference{} Perrin, M.\ 2007, Ph.D.\ Thesis, University of California, 
Berkeley

\reference{} Pooley, D., et al.\ 2002, ApJ, 572, 932

\reference{} Predehl, P., \& Schmitt, J.~H.~M.~M.\ 1995, A\&A, 293,
889

\reference{} Prieto, J. L., Garnavich, P., Chronister, A., \& Connick,
P.\ 2006, CBET, 648, 1

\reference{} Quimby, R.\ 2006a, PhD Thesis, University of Texas at Austin

\reference{} Quimby, R.\ 2006b, CBET, 644, 1

\reference{} Quimby, R., Castro, F., Mondol, P., Caldwell, J., \&
Terrazas, E.\ 2007, CBET, 793, 1

\reference{} Riess, A. G., et al.\ 1999, \aj, 118, 2675

\reference{} Richmond, M. W., et al.\ 1996, AJ, 111, 327

\reference{} Salamanca, I., Cid-Fernandes, R., Tenorio-Tagle, G.,
Telles, E., Terlevich, R.J., \& Munoz-Tunon, C.\ 1998, MNRAS, 300, L17

\reference{} Salamanca, I., Terlevich, R.J., \& Tenorio-Tagle, G.\
2002, MNRAS, 330, 844

\reference{} Scannapieco, E., Madau, P., Woosley, S. E., Heger, A., \&
Ferrara, A.\ 2005, ApJ, 633, 1031

\reference{} Schlegel, E. M.\ 1990, MNRAS, 244, 269

\reference{} Schlegel, E. M., \& Petre, R.\ 2006, ApJ, 646, 378

\reference{} Schlegel, D. J., Finkbeiner, D. P., \& Davis, M., 1998,
ApJ, 500, 525

\reference{} Smith, N., 2006, ApJ, 644, 1151

\reference{} Smith, N., 2007, AJ, 133, 1034

\reference{} Smith, N., \& Morse, J. A., 2004, ApJ, 605, 854

\reference{} Smith, N., \& Owocki, S. P.\ 2006, ApJ, 645, L45

\reference{} Smith, N., Vink, J.S., \& de Koter, A.\ 2004, ApJ, 615, 475

\reference{} Smith, N., et al.\ 2003, AJ, 125, 1458

\reference{} Sollerman, J., Cumming, R. J., \& Lundqvist, P.\ 1998,
\apj, 493, 933

\reference{} Stathakis, R. A., \& Sadler, E. M. 1991, MNRAS, 250, 786

\reference{} Sutherland, P. G., \& Wheeler, J. C.\ 1984, ApJ, 280, 282

\reference{}Taam, R. E., \& Ricker, P. M. 2006, astro-ph/0611043

\reference{} Turatto, M., et al. 1993, MNRAS, 262, 128

\reference{} Van Dyk, S. D.\ 2005, in The Fate of the Most Massive
    Stars, ed. R. Humphreys \& K. Stanek (San Francisco: ASP), 47

\reference{} Van Dyk, S. D., Weiler, K. W., Sramek, R. A., \& Panagio,
N.\ 1993, ApJ, 419, L69

\reference{} Van Dyk, S. D., Filippenko, A. V., \& Li, W.\ 2002, PASP,
114, 700

\reference{} Williams, C. L., Panagio, N., Van Dyk, S. D., Lacey, C. K.,
Weiler, K. W., \& Sramek, R. A.\ 2002, ApJ, 581, 396

\reference{} Woosley, S. E., \& Heger, A.\ 2006, ApJ, 637, 914

\reference{} Yoon, S. C., \& Langer, N.\ 2005, A\&A, 443, 643


\end{references}
\end{document}